\documentclass{article}
\usepackage{spconf,amsmath,graphicx}
\usepackage{hyperref}

\usepackage{booktabs}
\usepackage{amssymb}
\usepackage{array}
\usepackage{float}
\usepackage{enumitem} 

\title{AS-Speech: Adaptive Style For Speech Synthesis}
\name{Zhipeng Li\textsuperscript{1,2}, Xiaofen Xing\textsuperscript{1}\sthanks{Corresponding author. }, Jun Wang\textsuperscript{2}, Shuaiqi Chen\textsuperscript{2}, Guoqiao Yu\textsuperscript{2}, Guanglu Wan\textsuperscript{2}, Xiangmin Xu\textsuperscript{1}}
\address{$^1$South China University of Technology, China, $^2$Meituan, China\\}
\begin{document}
%
\maketitle
\begin{abstract}
In recent years, there has been significant progress in Text-to-Speech (TTS) synthesis technology, enabling the high-quality synthesis of voices in common scenarios. In unseen situations, adaptive TTS requires a strong generalization capability for speaker style characteristics. However, the existing adaptive methods can only extract and integrate coarse-grained timbre or mixed rhythm attributes separately. In this paper, we propose AS-Speech, an adaptive style methodology that integrates the speaker timbre characteristics and rhythmic attributes into a unified framework for text-to-speech synthesis. Specifically, AS-Speech can accurately simulate style characteristics through fine-grained text-based timbre features and global rhythm information, and achieve high-fidelity speech synthesis through the diffusion model. Experiments show that our proposed model produces voices with higher similarity in terms of timbre and rhythm compared to a series of adaptive TTS models while maintaining the naturalness of synthetic speech. Samples are available at \href{https://leezp99.github.io/as-speech-demo/}{https://leezp99.github.io/as-speech-demo/}
\end{abstract}
\vspace{-0.8em}
\section{Introduction}
\label{sec:intro}

In recent years, with the development of generative modeling\cite{ho2020denoising,vaswani2017attention}, non-autoregressive acoustic models\cite{ren2020fastspeech,shen2023naturalspeech}, and the efficient vocoder\cite{kong2020hifi,lee2022bigvgan}, Text-to-Speech (TTS) synthesis models have shown outstanding performance. Non-autoregressive models enjoy better robustness and generation speed due to predicting the features explicitly and simultaneously. Generative models like diffusion\cite{liu2022diffsinger, popov2021grad, kim2022guided}, flow\cite{kim2021conditional,kong23_interspeech}, etc., ensure the quality and diversity of generated voices. With the emergence of numerous applications like voice assistants, TTS's objectives have progressed from synthesizing speech for a single speaker to generating high-quality speech for multiple speakers and further advancing to support personalized voices\cite{chen2020adaspeech,min2021meta,casanova2022yourtts,kang2023grad,yoon2023sc}. This requires TTS models to generate high-quality speech while also adaptively and accurately capturing the speaking style of a given target segment, including characteristics both timbre and rhythm.

In the current research on rhythm in speech styles, maintaining a high consistency between the rhythm of speech and the generated text's overall semantic content is crucial. It ensures the production of speech with high naturalness and credibility. Therefore, prioritizing global rhythmic features over fine-grained features aligns more closely with practical needs. Considering the strong correlation between rhythm and emotion, adaptive rhythm methods can be highly similar to adaptive emotion methods. EmoMix\cite{tang23_interspeech} utilizes a pre-trained emotion Encoder to synthesize emotional rhythmic speech. CSEDT\cite{li2022cross} utilizes gradient reversal and orthogonalization to separate emotional information and adds it to the textual representation to fuse emotional rhythm. Appropriately combining rhythmic elements enhances the accuracy and naturalness of emotion expressive in speech synthesis systems. The research above convincingly demonstrates that global emotion features are sufficient to express emotional rhythm in speech. However, in adaptive text-to-speech, only considering emotional rhythmic factors is insufficient. Previous adaptive emotion TTS models only synthesize emotional speech in the seen speaker voices, which poses significant limitations in real-world scenarios. In practice, we need to consider not only emotional rhythm but also speaker timbre information.  

\begin{figure*}[ht]

    \centering
    \includegraphics[width=18cm]{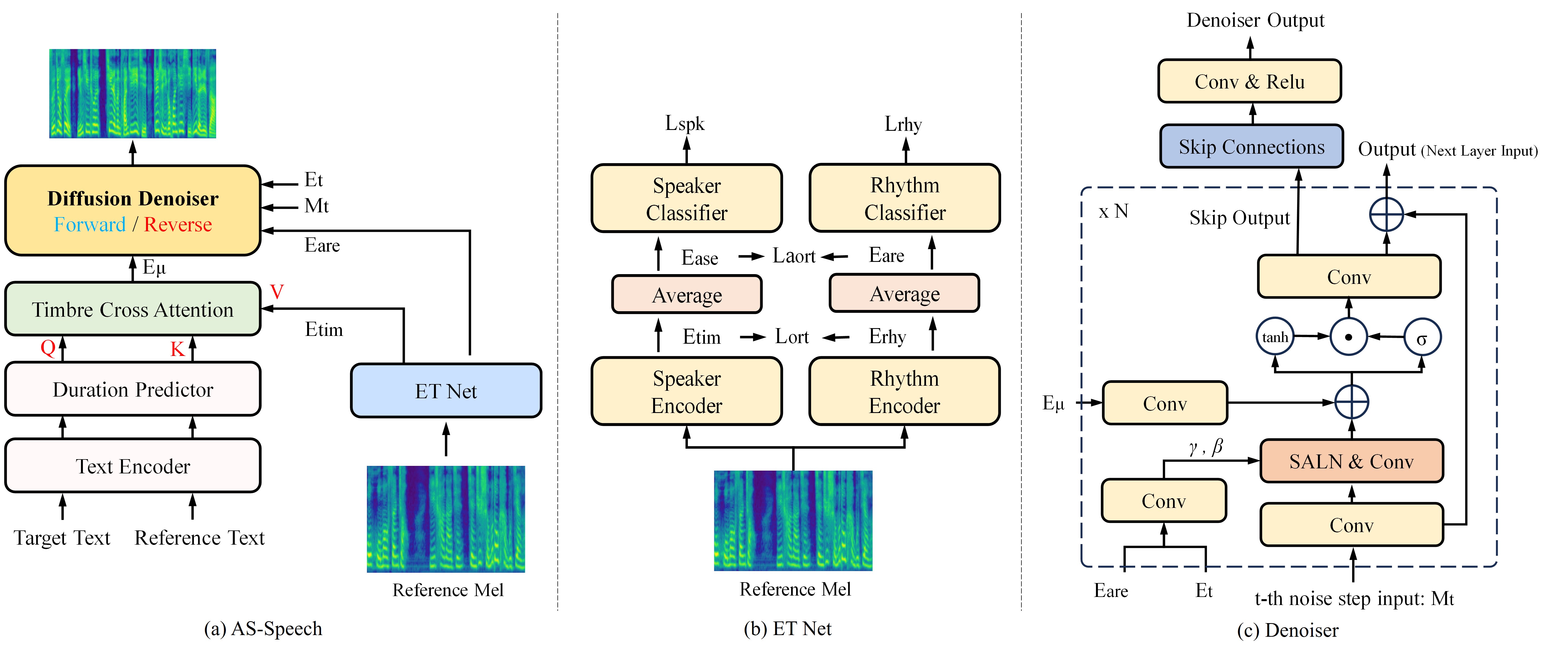}
    \caption{The overall architecture of the proposed AS-Speech. }
    \label{fig:galaxy}
\end{figure*}

Currently, the predominant approach for adaptive timbre (Aka. zero shot) TTS models involves global speaker vectors or pre-trained speaker encoder\cite{desplanques20_interspeech}. For instance, YourTTS utilizes a pre-trained speaker encoder\cite{heo2020clova} and introduces speaker consistency loss to enhance the similarity with the target segment's timbre. Adaspeech4\cite{wu22f_interspeech} tries to adapt the pre-trained model to the target speaker. Similarly, Meta-StyleSpeech employs global speaker vectors, embedding speaker attributes in a SALN (Style-Adaptive Layer Norm) manner. Likewise, Grad-StyleSpeech utilizes a mel-style encoder to extract average speaker vectors. However, employing global speaker vectors or pre-trained Speaker Encoders is a common practice but not optimal. By averaging speaker features over time, such methods result in a substantial loss of speaker timbre information, fail to support high-quality zero-shot speech synthesis with the target speaker's speaking styles and pronunciation habits. In this context, Attentron\cite{choi20c_interspeech} introduced a fine-grained encoder with a (text-audio) attention mechanism to extract styles from diverse reference samples. However, the relationship between text and speech is one-to-many, and the same word in text can correspond to a wide variety of pronunciations and different durations. It will be difficult for the cross-modal attention mechanism to capture the speaker information relationship between text and audio accurately, due to the natural differences and inconsistencies between text and audio domains. And the previous text-to-audio attention methods are not as good as the text-to-text attention method, which is more natural and interpretable.

Based on the above mentioned considerations, we propose AS-Speech, an adaptive style (both timbre and rhythm) methodology for text-to-speech synthesis. In contrast to previous adaptive approaches, AS-Speech can accurately simulate style characteristics through fine-grained timbre features based on text and global rhythm attributes according to a few seconds of reference segment, and achieve high-fidelity speech synthesis through the diffusion model.

Our main contributions can be summarized as follows:
\begin{itemize}
\item In this paper, we propose a style-adaptive TTS model named AS-Speech, which integrates the speaker timbre characteristics and rhythmic features inherent in style into a unified framework, can effectively produce style speech from the reference segment.
\item We propose a fine-grained timbre module based on text designed to extract and transfer the speaker local timbre features effectively and accurately from reference segment.
\item AS-Speech outperforms recent timbre adaptive models and rhythm adaptive models in generating stylized speech, as evidenced by the results of Style60 and VCTK datasets.
\end{itemize}

\section{Methodology}
\label{sec:method}

Adaptive Text-To-Speech (TTS) aims to synthesize the speech given target text transcription ${{X}}_{t}$ and reference mel-spectrogram ${{M}}_{r}$ of the target speaker. Unlike the previous approaches, we also employ the text ${{X}}_{r}$ of reference segments to enhance the precise capture of the speaker's fine-grained timbre features.
The overall architecture of the proposed AS-Speech is shown in Figure 1. Our model consists of five main parts: text encoder, duration predictor, ET net, timbre cross-attention module (TCA), and diffusion module. The text encoder adopts stacked transformer blocks, and the duration predictor is based on NAT\cite{shen2020non}. ET network is designed for learning timbre and rhythm features at different granularities under label and orthogonality constraints. The timbre cross-attention module captures fine-grained timbral information embedded in reference speech by leveraging pronunciation similarity relationships between the target text and reference text. To effectively incorporate global rhythmic features, we employ a modified WaveNet\cite{oord2016wavenet} as the underlying denoiser network. Details about these components are presented in the following sections.

\subsection{ET Net}

Mel-spectrogram conveys a rich information stream, containing content, timbre, rhythm, and other components. Directly using mel-spectrogram for adaptive TTS may result in suboptimal performance. Hence, it is crucial to maximize the extraction of pure timbre and rhythm features. In ET Net, we employ label supervision and multi-granularity orthogonal loss to disentangle speaker identity and rhythmic features from the reference spectrogram. We use reference spectrogram from another speech of the same speaker for each training text-speech example. 

ET Net takes ${{M}}_{r}\in\mathbb{R}^{80 \times T_{r}}$ as input, where $\textit{T}_{r}$ is the number of reference mel-spectrogram frames. ${{M}}_{r}$ is fed into each encoder to get fine-grained timbre and rhythm embeddings, denoted as ${{E}}_{tim},{{E}}_{rhy}\in\mathbb{R}^{F \times T_{r}}$, $\textit{F}$ indicates the feature dims. Subsequently, the timbre and rhythm representation go through the average pooling layers over time dimension to obtain average speaker and rhythm embedding, denoted as ${{E}}_{ase},{{E}}_{are}$. Then, we introduce classifiers to predict the speaker and rhythm label of ${{E}}_{ase},{{E}}_{are}$ separately, and use supervision for timbre and  rhythm to achieve high-quality speech disentanglement. $E_{tim}$ and $E_{rhy}$ are separately fed into the Timbre Cross-Attention module and Diffusion Module. We employ supervised learning ${{L}}_{spk}$ with speaker labels and ${{L}}_{rhy}$ rhythm labels to ensure each encoder working correctly.

To make two global embeddings unrelated, CSEDT \cite{li2022cross} proposes orthogonality loss. Unlike simply minimizing orthogonality loss ${{L}}_{aort}$ at a coarse granularity, the simultaneous consideration of fine-grained orthogonality loss ${{L}}_{ort}$ aids in a more comprehensive decoupling and preservation of timbre and rhythm information.

\begin{gather}
    L_{aort}=||E_{ase} \cdot E_{are}||_F^2 \\
    L_{ort}=||\sum_{i=1}^{T_{r}} E_{tim}^i \cdot E_{rhy}^i||_F^2
\end{gather}

where $||\cdot||_F$ is the Frobenius norm, $E_{tim}^i, E_{rhy}^i$ is the i-th frame of $E_{tim}, E_{rhy}$.

Label-supervised learning ensures that vectors derived from ${{M}}_{r}$ acquire as many specific properties as possible, such as timbre or rhythm, whereas orthogonal minimization learning lets them be unrelated, resulting in purer feature properties, which is beneficial for more effective control over the transfer of timbre and rhythm.

\subsection{Timbre Cross-Attention Module}
Pervious zero-shot studies typically employ universal speaker embeddings derived from reference audio. Those approaches neglect the transmission of individual phonetic attributes linked to phoneme content, resulting in poor speaker likeness with respect to detailed speaking styles and pronunciation patterns. For neutral speech, speakers exhibit highly similar or even identical pronunciations of the same word or phoneme. To enhance the similarity in speaker pronunciation between synthesized speech and the reference, we introduce a module that leverages the content relationship between the target text Xt and the reference text Xr to guide local pronunciation transfer, we called this module as Text-based Timbre Cross-Attention Module (TCA).

\begin{figure}[t]
  \centering
  \includegraphics[width=9cm]{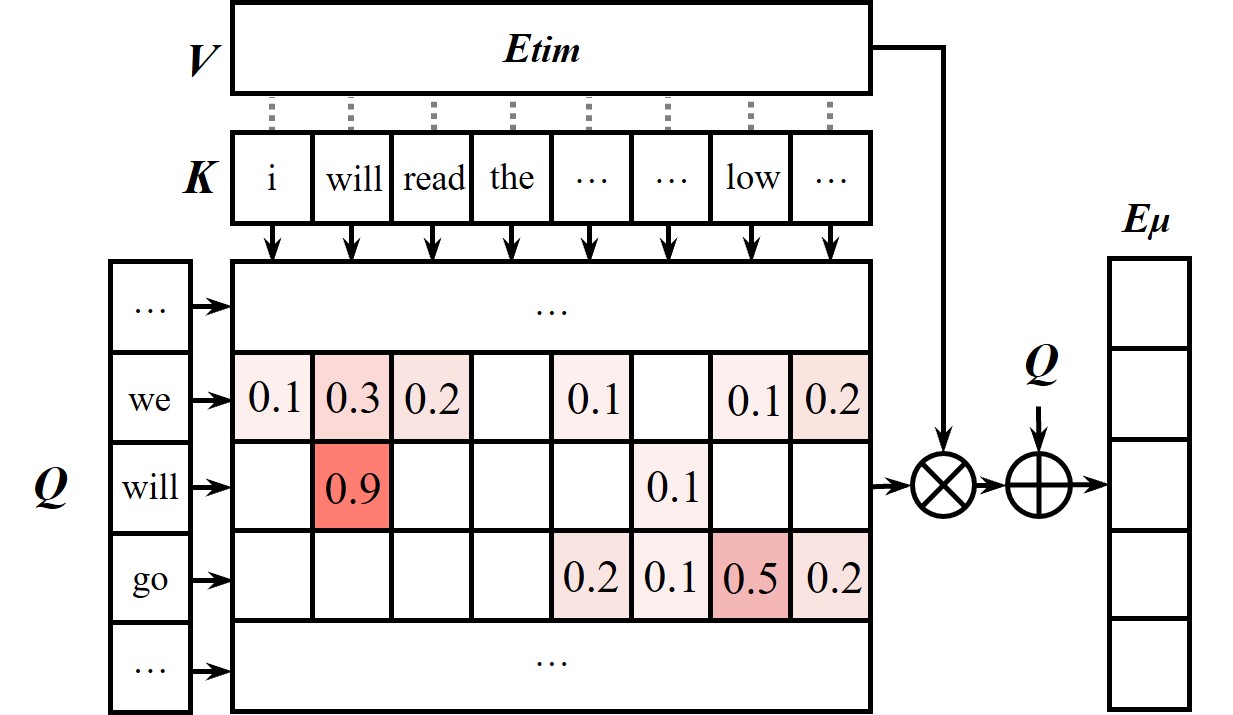}
  \caption{The details of TCA module. Attention scores reflect the phonetic similarity of characters from Query and Key.}
  \label{fig:narrow_image}
  \vspace{-1.0em}
\end{figure}

Target text ${{X}}_{t}$ and reference text ${{X}}_{r}$ are processed through a Text Encoder and Duration Predictor to obtain frame-level text representations, denoted as ${{X}}_{t}^{'}, {{X}}_{r}^{'}$. The frame-level text representation ${{X}}_{r}^{'}$ and fine-grained timbre embedding ${{E}}_{tim}$ are temporally correspondent. Subsequently, in TCA, target text representation is Query, reference text representation is Key, and the fine-grained timbre feature ${{E}}_{tim}$ acts as Value. As shown in Figure 2, the content attention matrix between ${{X}}_{t}^{'}$ and ${{X}}_{r}^{'}$ guides the selection of fine-grained timbre representations, ensuring a high degree of similarity in pronunciation for the same phoneme. Query residual connections are employed to ensure gradient stability. $E_\mu$ is fed to diffusion denoiser.

\vspace{-1.0em}
\begin{gather}
\label{attention}
E_{\mu}=TCA(Q,K,V)=softmax(\frac{QK}{\sqrt{F}})V+Q
\end{gather}

\subsection{Diffusion Module}
Diffusion is a generative model built upon a forward process fixed to a Markov chain that diffuses data $\textit{x}_0$ into white noise and a reverse process that generates samples $\textit{x}_0$ by progressive denoising the noise sampled from the prior noise distribution. The complete proof of formulas can be found in \cite{ho2020denoising,liu2022diffsinger}.

In recent years, diffusion models\cite{rombach2022high, zhang2023adding, choi23c_interspeech} have demonstrated outstanding performance in conditional generation.
As Figure 1(c) illustrates, AS-Speech employs a modified WaveNet with Style-Adaptive Layer Norm (SALN) as the underlying denoiser network $\theta$. We determine the scale $\gamma$ and bias $\beta$ with conditional inputs, the global rhythm embedding $E_{are}$ and the time embedding $E_{t}$.

\vspace{-1.0em}
\begin{gather}
SALN(X, \gamma, \beta)= \gamma * \frac{X-mean}{var} + \beta
\end{gather}

In the forward procedure, diffusion module takes in the mel-spectrogram at t-th noise step $M_t$ ($M_0$ means GT mel-spectrogram),$E_{\mu},E_{t}$, and $E_{are}$. Then, update the denoiser $\theta$ outputs $\epsilon_{\theta}(M_{t}, E_{\mu}, E_{t}, E_{are})$ by the gradient
in Formula 5.

\vspace{-0.3em}
\begin{gather}
\nabla_{\theta} \lVert \epsilon - \epsilon_{\theta}(\sqrt{\overline{\alpha}_{t}}M_{0}+\sqrt{1-\overline{\alpha}_{t}}\epsilon,E_{\mu}, E_{t}, E_{are}) \lVert^2
\end{gather}

The reverse procedure starts at the Guassian white noise $M_{T}$ sampled from $\mathcal{N}(0, I)$. Then the reverse diffusion iterates for $T$ times to predict the denoiser output $\epsilon_{\theta}$ and obtain $M_{t-1}$ from $M_{t}$ according to Formula 6, where $\textbf{z} \sim \mathcal{N}(0, I)$ except for $\textbf{z}=0$ when $t=1$.

\vspace{-1em}
\begin{gather}
M_{t-1}=\frac{1}{\sqrt{\alpha_t}}(M_t-\frac{1-\alpha_t}{\sqrt{1-\overline{\alpha}_{t}}}\epsilon_{\theta}(M_{t}, E_{\mu}, E_{t}, E_{are}))+ \sigma_t\textbf{z}
\end{gather}

Finally, a mel-spectrogram $M$ corresponding to $E_{\mu}, E_{are}$ could be generated.

\section{Experiments}
\label{sec:experiments}
\vspace{-1em}

\subsection{Dataset}
\noindent \textbf{Style60}: Style60 is a Mandarin Chinese style speech corpus collected internally. The corpus contains 23 hours of speech data from 60 speakers, and all the data are divided into eight rhythm categories, i.e., neutral, happy, angry, sad, afraid, news, story, and poetry. We have divided the Style60 dataset into the train set with 54 speakers and the test set with the remaining 6 speakers.

\noindent \textbf{VCTK}\cite{veaux2016superseded}: The dataset contains 44 hours of neutral speech data uttered by 109 native speakers of English with various accents. The train and test set of VCTK was used the same way as in previous studies\cite{casanova2022yourtts}.

\subsection{Experimental Setup}
We stack 8 transformer blocks for text encoder and 20 modified wavenet layers for diffusion denoiser. The feature dim $\textit{F}$ is 256, and the denoising steps are 100. To balance the weights of different losses, we set a scaling factor of 0.01 for $L_{ort}$ additionally. Our models are trained with 1M steps with batch size 16 on a single A100 GPU to ensure complete convergence. We employ pretrained universal HiFI-GAN vocoder for waveform generation.

\subsection{Evaluation Setup}
\subsubsection{adaptive rhythm experiments}
For rhythm-adaptive experiments, we conducted comparisons using the Style60 test set and introduced five metrics. The first is objective metric, Rhythm Classifier Accuracy (RCA), which measures the rhythm category of the synthesized speech. Specifically, we employ a pretrained rhythm classification model and predict the rhythm class of the synthesized speech. For subjective metrics, we use Mean Opinion Score (MOS) to measure the naturalness and rhythm Similarity MOS (R-SMOS) to measure the rhythm similarity of synthesized and reference speech. The remain two metrics are Speaker Embedding Cosine Similarity (SECS) and speaker Similarity MOS (S-SMOS).

Following, we provide detailed evaluation setups. (1) GT (voc.): speech generated from ground truth mel-spectrogram using HiFi-GAN. (2) GradTTS: GradTTS is an effective acoustic model based on diffusion, and we train it with hard rhythm labels as inputs. (3) CSEDT*: CSEDT\cite{li2022cross} is a cross-speaker emotion transfer method, and we adapt its major implementation to FastSpeech2 with rhythm labels, named CSEDT*. (3) AS-Speech (w/o ort): AS-Speech trained without the fine-grained orthogonal loss $L_{ort}$ (4) AS-Speech: this is our proposed model with ET Net, TCA, diffusion module, and trained with orthogonal losses, both $L_{aort}$ and $L_{ort}$. GradTTS and CSEDT* did not have adaptive timbre ability, so they only evaluated rhythm-related metrics.

\subsubsection{adaptive timbre experiments}
For speaker zero-shot experiments, we conducted comparisons using the VCTK dataset and introduced three metrics. The first is objective metric, Speaker Embedding Cosine Similarity (SECS). We measure the similarity between vectors of the synthesized and reference speech using the speaker encoder from the resemblyzer\cite{wan2018generalized} repository. For subjective metrics, we also use MOS to measure the naturalness of the synthesized speech and speaker Similarity MOS (S-SMOS) to measure the speaker similarity of synthesized and reference speech.

\begin{table*}
    \centering
    \setlength{\extrarowheight}{4pt}
    \begin{tabular}{l@{\hspace{12mm}}c@{\hspace{10mm}}c@{\hspace{10mm}}c@{\hspace{10mm}}c@{\hspace{10mm}}c@{\hspace{10mm}}}
        \bottomrule
        \textbf{Model} & MOS ($\uparrow$) & RCA ($\uparrow$) & R-SMOS ($\uparrow$) & SECS ($\uparrow$) & S-SMOS ($\uparrow$) \\
        \hline
        \textbf{GT (voc.)} & $4.413_{\pm0.039}$ & 63.6\% & $4.122_{\pm0.040}$ & $87.23_{\pm1.00}$ & $4.197_{\pm0.048}$ \\
        \hline
        \textbf{GradTTS} & $3.641_{\pm0.052}$ & 56.0\% & $3.658_{\pm0.068}$ & --- & --- \\
        \textbf{CSEDT*} & $3.901_{\pm0.052}$ & 62.3\% & $3.990_{\pm0.052}$ & --- & --- \\
        \hline
        \textbf{AS-Speech (w/o $L_{ort}$)}& $4.289_{\pm0.041}$ & 65.6\%  & $4.034_{\pm0.039}$ & $81.73_{\pm1.13}$ & $3.637_{\pm0.070}$ \\
        \hline
        \textbf{AS-Speech} & $\textbf{4.349}_{\pm\textbf{0.039}}$ & \textbf{66.3}\% & $\textbf{4.075}_{\pm\textbf{0.039}}$ & $\textbf{83.16}_{\pm\textbf{1.06}}$ & $\textbf{3.650}_{\pm\textbf{0.069}}$ \\
        \bottomrule
    \end{tabular}
    \caption{Adaptive experimental results on Style60 test set with confidence intervals of 95\% (except for RCA).}
    \label{tab:my_label}
\end{table*}

Details for each method we used are described as follows: 
(1) GT (voc.) (2) StyleSpeech: an adaptive speaker TTS model using a learnable mel encoder. (3) YourTTS: a zero-shot TTS model with a fixed speaker encoder. (4) AS\_xvector: we employ the global speaker embedding extracted by pretrained ECAPA-TDNN\cite{desplanques20_interspeech, ravanelli2021speechbrain} to the AS-Speech* backbone (w/o ET net and TCA). (5) AS\_ase: we use AS-Speech* backbone (w/o TCA) and employ $E_{ase}$ instead of $E_{tim}$, simply adding it to the output of text encoder. (6) AS-Speech*: we removed the rhythm-related components of AS-Speech, called AS-Speech*, due to VCTK being a single-style dataset. 
For calculation of SECS, MOS, and S-SMOS in English, we follow the same sentences of YourTTS (sentences are chosen in LibriTTS\cite{zen2019libritts} dataset with more than 20 words). 
So, The MOS evaluation of GT is only reported, and samples are randomly chosen from the VCTK test set. 
This experiment aims to compare the adaptive speaker capabilities of various methods.

\begin{table}
    \setlength{\extrarowheight}{4pt}
    \begin{tabular}{lccc}
        \bottomrule
        \textbf{Model} & MOS ($\uparrow$) & SECS ($\uparrow$) & S-SMOS ($\uparrow$) \\
        \hline
        \textbf{GT (voc.)} & $4.053_{\pm0.048}$ & --- & --- \\
        \hline
        \textbf{StyleSpeech} & $3.424_{\pm0.053}$ & $84.66_{\pm1.10}$ & $3.368_{\pm0.068}$\\
        \textbf{YourTTS} & $3.899_{\pm0.049}$ & $86.09_{\pm0.89}$ & $3.793_{\pm0.061}$\\
        \hline
        \textbf{AS\_xvector} & $3.793_{\pm0.047}$ & $84.33_{\pm1.59}$ & $3.419_{\pm0.086}$\\
        \textbf{AS\_ase} & $\textbf{3.968}_{\pm\textbf{0.050}}$ & $82.19_{\pm1.19}$ & $3.182_{\pm0.096}$\\
        \hline
        \textbf{AS-Speech*} & $3.931_{\pm0.047}$ & $\textbf{87.30}_{\pm\textbf{1.08}}$ & $\textbf{4.007}_{\pm\textbf{0.055}}$\\
        \bottomrule
    \end{tabular}
    \caption{Evaluation results for zero-shot timbre adaptation on VCTK test set.}
    \label{tab:my_label}
\end{table}

\subsection{Experimental Results}
\subsubsection{results analysis}
We now validate the adaptation performance of our model on the Style60 (Mandarin) test set. To this end, we first evaluate the quality of generated speech. In Table 1, the results of MOS show that \textbf{AS-Speech} achieves the best generation quality, largely outperforming the baselines (\textbf{GradTTS}, \textbf{CSEDT*}), which demonstrates that our backbone is an excellent acoustic model. For the rhythm adaptive experiments, results show that our AS-Speech beats other adaptive methods, even ground truth in terms of RCA, shows that AS-Speech is able to effectively extract the rhythm attributes and synthesize the style speech conditioned on the reference speech's rhythm. In the subjective evaluation of R-SMOS, the proposed method also reaches higher scores than other models, +0.417 to \textbf{GradTTS} and +0.085 to \textbf{CSEDT*}, proves that the synthesized speech's rhythm from our approach more closely resembles reference speech's rhythm compared to prior methods, and this is attributed to the incorporation of the SALN module within the diffusion model. As the diffusion denoiser processing from timestep T-1 to 0, the global rhythm representation can be fully integrated into the generated mel-spectrograms. 
The performance of \textbf{GradTTS} trained with hard labels is subpar, potentially due to the limitation of a discrete one-hot vector representing rhythm categories, which may not adequately capture subtle and rich rhythm variations, leading to a lack of diversity. So adaptive style model should catch rhythm presentation from the reference speech rather than setting hard rhythm label.

We conduct ablation studies to verify the effectiveness of fine-grained orthogonality loss. After training with fine-grained orthogonal loss, a slight improvement was observed in both speaker and rhythm similarity metrics. This indirectly indicates that the speech rhythm is intertwined with timbre, and obtaining a pure rhythm representation can enhance adaptive style ability. It is worth mentioning that this only requires addition during training and does not affect any inference speed.

Shown in Table 2, we present evaluation results for zero-shot adaptation performance on VCTK test set (unseen speakers). In short, our model outperforms other methods, whether in objective or subjective evaluations.

\begin{figure}[t]
  \centering
  \includegraphics[width=9cm]{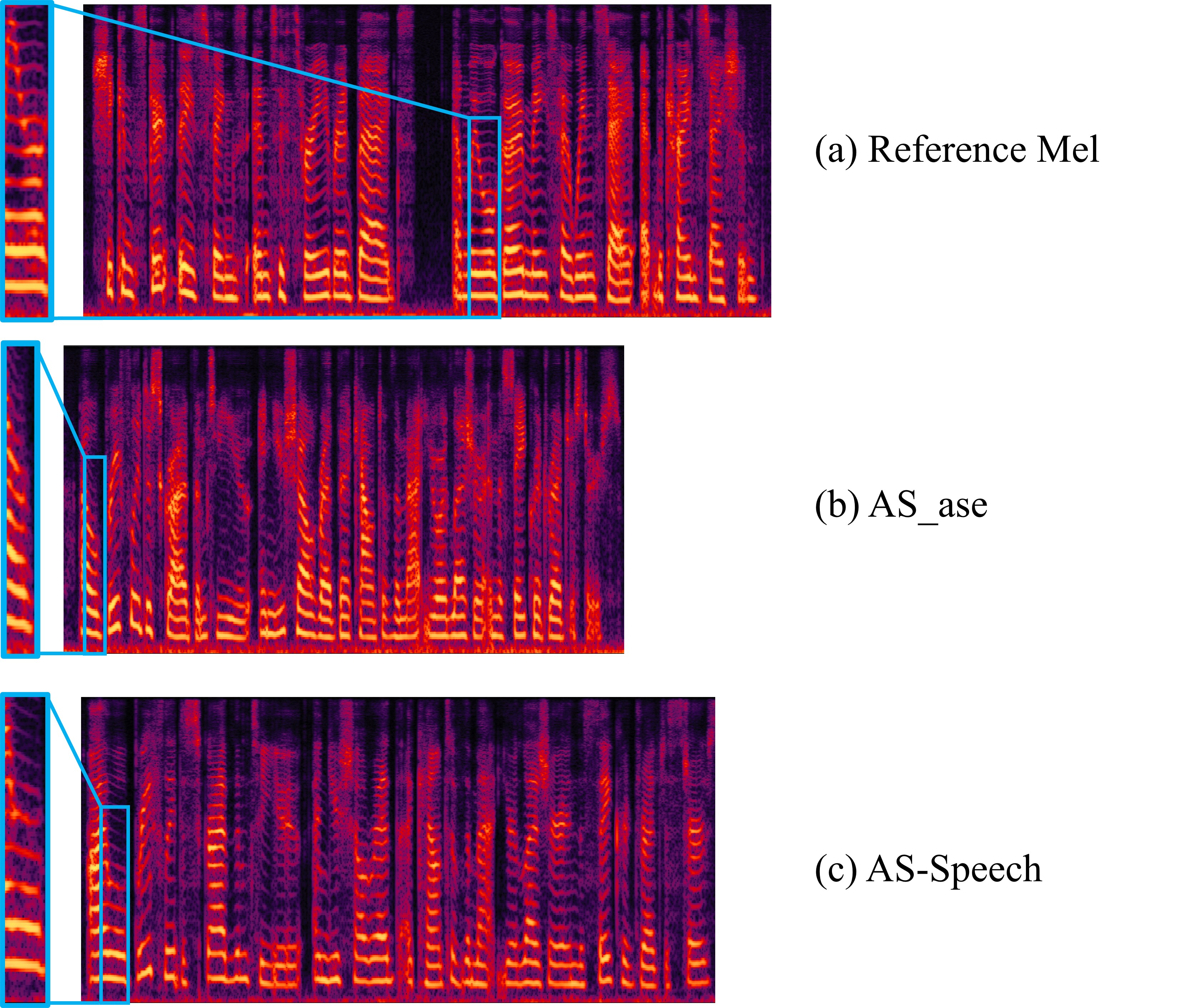}
  \caption{Visualizations of synthesized and reference mel-spectrograms. Target text and reference text both have the word "will" (blue box) }
  \label{fig:narrow_image}
\end{figure}

In the ablation experiments conducted on \textbf{AS\_xvector} and \textbf{AS-Speech*}, the improvement on SMOS, SECS is significant by a gap of over 0.6, 3, indicating that a learnable speaker encoder holds greater potential compared to a fixed, general encoder. In another ablation experiments, the SECS and S-SMOS scores obtained by \textbf{AS\_{ase}} and \textbf{AS-Speech*} indicate that fined-grained timbre representation contains more speaker information than global speaker embeddings and the TCA module based on text are helpful to improve the speaker similarity from the sense of listening for zero-shot speaker adaptation. 

On naturalness, \textbf{AS-Speech*} surpasses other models on MOS results, matching \textbf{AS\_{ase}}, due to the strong generative capability of Diffusion.

\subsubsection{visualization analysis}
To further demonstrate the effectiveness of TCA module, we plot the mel-spectrograms from \textbf{AS-Speech*}, \textbf{AS\_{ase}}, and reference speech in Figure 3, The reference audio’s text and target text both include the word ``will”, and the blue box represents the pronunciation region of the word ``will" (phoneme is ``W AH0 L"). 
We compared the mid-low frequency areas of Mel corresponding to the word “Will” and observed that the resonance peak trends in Figure 3c are very similar to Figure 3a, significantly surpassing those in Figure 3b.  This demonstrates that TCA can extract speaker’s local prounication habits and successfully transfers the reference speaker’s timbre based on text similarity. The any-speaker adaption performance is derived from TCA module and fine-grained timbre representation, and that aligning perfectly with our design intent.

The above experiments demonstrate that \textbf{AS-Speech} can synthesize style speech based on the provided reference speech, effectively integrating timbre adaptation and rhythm adaptation into one acoustic model.

\section{Conclusion}
In this work, we have proposed AS-Speech, a style-adaptive TTS model that integrates timbre and rhythm representations into a unified framework and can accurately simulate target style characteristics according to a few seconds of speech. Our approach employs ET net to obtain fine-grained speaker information and speaker-irrelevant rhythm embedding. And the timbre cross-attention module based on text can extract and transfer speaker timbre features effectively. We utilize a conditional diffusion module with SALN to generate the high-fidelity style speech. The experiment results on Style60 and VCTK show that the quality of generated speech from AS-Speech highly outperforms previous adaptive methods in objective and subjective measures of both timbre and rhythm.

\bibliographystyle{IEEEbib}
\bibliography{strings,refs}

\end{document}